# Tourism destination events classifier based on artificial intelligence techniques

**Miguel Camacho-Ruiz[1], Ramón Alberto Carrasco[2], Gema Fernández-Avilés[3], Antonio LaTorre[4]**

[1] Faculty of Commerce and Tourism Complutense, University of Madrid, 28223 Madrid, Spain; mcamacho@atalayatech.com.

[2] Department of Marketing, Faculty of Statistics, Complutense, University of Madrid, 28040 Madrid, Spain; ramoncar@ucm.es

[3] Faculty of Legal and Social Sciences, Toledo, University of Castilla—La Mancha, Cobertizo de San Pedro Mártir, S/N. 45.071, Toledo, Spain; Gema.FAviles@uclm.es. Corresponding author.

[4] Center for Computational Simulation, Universidad Politécnica de Madrid, 28660, Madrid, Spain; a.latorre@upm.es

## Abstract:

Identifying client needs to provide optimal services is crucial in tourist destination management. The events held in tourist destinations may help to meet those needs and thus contribute to tourist satisfaction. As with product management, the creation of hierarchical catalogs to classify those events can aid event management. The events that can be found on the internet are listed in dispersed, heterogeneous sources, which makes direct classification a difficult, time-consuming task. The main aim of this work is to create a novel process for automatically classifying an eclectic variety of tourist events using a hierarchical taxonomy, which can be applied to support tourist destination management. Leveraging data science methods such as CRISP-DM, supervised machine learning, and natural language processing techniques, the automatic classification process proposed here allows the creation of a normalized catalog across very different geographical regions. Therefore, we can build catalogs with consistent filters, allowing users to find events regardless of the event categories assigned at source, if any. This is very valuable for companies that offer this kind of information across multiple regions, such as airlines, travel agencies or hotel chains. Ultimately, this tool has the potential to revolutionize the way companies and end users interact with tourist events information.

## 1. Introduction

Knowing what events are available in a destination can be an important factor in a tourist's decision-making process. Events can enhance the attractiveness of a destination and provide unique experiences for tourists [1]. They can boost tourist traffic, encourage tourism spending, and build the identity of the tourist destination [2]. Tourists may be attracted to destinations that host events related to their interests, such as music festivals, sports competitions, cultural festivals, or conferences. Classifying an event can help in the travel decision process by making it easier for tourists to find and choose events that align with their interests and preferences.

From the perspective of tourism destination management, it is crucial to know clients' needs in order to be able to adapt services accordingly. Tourist events linked to a destination are an important part of these services. Such events are an important driver of tourism, and figure prominently in the development and marketing plans of most destinations. In order to develop the potential of tourist destinations, event managers should be involved in the planning process [3]. However, tourism is a fragmented and disjointed activity [4,5,6] and we do not see the mass adoption of event segmentation. Most tourist events are publicized on the internet via aggregators, affiliated players, or organizer sites. While they are usually categorized, there is not a widely-used standard for event taxonomies. Indeed, there are only a few fields that are used extensively: title/name, description, and place. Moreover, none of those fields are structured or normalized. Having a client-centered tourism product taxonomy is vital to be able to create a marketing-oriented product category system. Some taxonomy systems have been proposed in previous works, such as [7] or [8]. They suggest that the tourism sector should adopt a standard taxonomy for tourist events, but encouraging widespread adoption is a challenge.

Despite the importance of event segmentation in tourism destination management, there is still a need for a standardized taxonomy to classify events. In seeking to address this gap, our goal is to pioneer a process that automatically classifies tourist events of diverse types, sourced from varied locations, using a hierarchical taxonomy. This innovative approach permits the creation of a normalized event catalog. Our solution accommodates the needs of travelers in the inspirational phase of their trip-planning, providing access to a consistent classification of events, regardless of source, language, or original category. By extension, our work also significantly benefits companies that offer tourism services in different regions.

By applying the process developed, we can create a live catalog of events with a normalized taxonomy, regardless of the category (if any) assigned by the event listing source. This is especially useful for travelers in the inspirational phase of their trip-planning, as they can use the same taxonomy regardless of the event source, language or original category. Since it is useful for the potential tourist, it is also useful for companies that offer tourism services in different regions or want to promote destinations, independently of event listing sources. Such companies may include airlines, OTAs, traditional travel agencies, hotel chains, etc.

The methodology used for this work is CRISP-DM, a European open standard process model that describes common approaches used by data mining experts [9]. In the context of CRISP-DM, other techniques are used, such as machine learning (ML)—specifically supervised learning—and natural language processing (NPL). The main aim is to apply our model to real-world data; therefore, in addition to the model itself, we propose an underlying architecture that allows its practical application.

To create a classifier for tourism events we need to use a statistically significant number of heterogeneous events to train our model. In this study, we used 1103 event listing sources from 30+ countries to collect 700,000 events listed in 23 languages. The remainder of the article is structured as follows: Section 2 discusses the state of the art and compares the studies related to our proposal; Section 3 presents the foundations of NPL and supervised learning, on which our proposal is based; Section 4 describes the model for classification; Section 5 shows the results obtained by this model and a use case; finally, Section 6, draws conclusions and suggests future avenues of work.

## 2. Literature review

Classic studies show that market segmentation can be a good management and marketing strategy [10]. Segmentation involves viewing a heterogeneous market as several smaller, more homogeneous markets [11]. In tourism, it can aid decision-making; indeed, segmentation based on tourist events is a hot topic in the industry, but little work has been done on how to apply a taxonomy model to current or future events.

There are many criteria to use when classifying events, and different areas of classification, such as thematic or geographical. In this section, we include only those papers that are directly related to general events and ways to taxonomize and classify them.

**Table 1. Studies on tourism event classification**

| Ref. | Fundamentals | Key features | Application | Classification Method |
|---|---|---|---|---|
| McKercher (2016) [8] | Phenetic method to produce a hierarchical taxonomy | 330 hierarchical categories in 5 top level categories. Those categories are indicative but not exhaustive. | Taxonomy proposal for tourism event classification | No classification method |
| Arcodia and Robb (2000) [12] | Compilation and review of many publications | Provides 5 categories for Australian event industry | Categorization of terminology in event management research and practice | No classification method |
| Getz (2016) [13] | Compilation and review of many publications | Many categories depending on different criteria | Taxonomy proposal for tourism event classification | No classification method |
| Gibson (2002) [15] | Compilation and review of many publications | 3 overlapping typologies | Typology proposal for sport events | N/A |
| Gjorgievski, Kozuharov, and Nakovski (2013) [14] | Compilation and review of many publications | Many categories depending on different criteria | Taxonomy proposal for tourism event classification | No classification method |
| Gammon (2020) [16] | Compilation and review of many publications | 5 typologies | Typology proposal for sport events | No classification method |
| Getz (2005) [17] | Compilation and review of many publications | 7 typologies | Typology proposal for 'event studies' | Suggests the creation of a lobby or marketing consortium |
| Cepeda-Pacheco & Domingo (2022) [18] | Recommendation system for tourist attractions in smart cities | It uses labels for event classification, but it does not propose a taxonomy system or suggest how to generate labels | Recommendation system to help tourists with decision-making. | No classification method for tourism events, but it uses Neural Networks to recommend events to tourists |
| Iliev (2020) [19] | Compilation and review of many publications | It reviews categorization and division of religious activities for tourists | Segmentation of religious events | No classification method |
| **Our proposal** | **Combination of taxonomies suggested by Getz (2005) and evolution of dataset used in private industry** | **8 main categories with variable number of subcategories (100+)** | **Classification of events using a modification of Getz taxonomy proposal** | **Using NLP, ML and big data techniques** |

As can be deduced from a review of the related literature, most papers propose taxonomies or typologies, but do not provide clear criteria for classifying events. Indeed, in most cases the researchers defer to the event organizer or event listing source when it comes to assigning a taxonomic category. Some papers even call for the formation of an event management lobby [17] to help with the task, while others propose a recommendation system with automatic segmentation of events based on user preferences [18]. However, we found no studies that propose a way to automatically classify events on the basis of the title and description of said event.

While the related literature highlights the importance of having labeled tourism events, none of the studies propose an automatic classification method; this presents an opportunity to propose a novel solution. Our work is innovative since it provides an automated, scalable, language- and source-agnostic classification method. To our knowledge, it is the first of its kind for tourism events.

## 3. Methodology

The proposed method for automatically classifying events relies on two main systems: a language model based on BERT (Bidirectional Encoder Representation from Transformers) and a logistic regression model for classification.

### *3.1 BERT*

BERT is currently the preeminent, state-of-the-art natural language model [20]. It uses a bidirectional transformer to train a language model on a large corpus originally provided by Google, and can be fine-tuned to be trained for other tasks provided by the final user. For classification purposes, BERT takes an array of tokens representing a sentence and a mask that helps it determine how long the sentence is. If fine-tuning is used for classification, BERT will learn and update its internal model at the same time as the classifier learns.

BERT has many variants depending on its configuration. In [21], two configurations are proposed for the transformer: namely, BERT-base and BERT-large. The one used in this work is BERT-base, a model with 12 transformer blocks, 768 hidden layers, 12 self-attention heads and 110 million parameters. BERT-large, a larger version of the model, was tested for this study but resulted in only marginally better results at the cost of far longer processing times.

BERT is composed of 6 identical layers. In turn, each of these layers is composed of a multi-head self-attention mechanism and a position-wise fully connected feed-forward network. It uses a residual connection [22] to connect both sub-layers followed by a layer normalization [23].

In relation to multi-head self-attention, first, we need to define scaled dot-product attention. It is defined as follows:

$$Attention(\boldsymbol{Q},\boldsymbol{K},\boldsymbol{V}) = softmax\left(\frac{\boldsymbol{Q}\boldsymbol{K}^T}{\sqrt{d_k}}\right)\boldsymbol{V}$$

where **Q** is the matrix of queries, **K** is the matrix of keys, **V** is the matrix of values and dk is the dimension of the **Q** and **K** matrices. Now, we can define multi-head attention as follows:

$$MultiHead(\boldsymbol{Q}, \boldsymbol{K}, \boldsymbol{V}) = Concat(head_1, \dots, head_2)W^{\boldsymbol{o}}$$
$$where\ head_i = Attention(\boldsymbol{Q}\boldsymbol{W}_i^{\boldsymbol{Q}}, \boldsymbol{K}\boldsymbol{W}_i^{\boldsymbol{K}}, \boldsymbol{V}\boldsymbol{W}_i^{\boldsymbol{V}})$$

Multi-head attention consists of projecting the queries, keys and values h times with different, learned linear projections to dk, dk and dv (dimension of the values matrix), respectively. Then, on each of these projected versions of the queries, keys and values, we perform the attention function in parallel, yielding $d_v$-dimensional output values. Finally, these are concatenated and projected, resulting in the final values [24].

BERT empirically outperforms traditional NLP approaches for text classification in different datasets [25]. BERT also outperforms classification algorithms such as bag-of-words algorithms for tasks like the classification of Yelp shopping reviews [26].

*3.2 Logistic regression*

Logistic regression is a kind of regression analysis used to predict the outcome of a categorical/discrete/binary variable based on the independent variables provided. It is a useful way of modeling the probability of an event happening as a function of other factors. Logistic regression analysis falls within the framework of Generalized Linear Models (GLM). It can be used in combination with some other ML techniques to classify non-binary categorical variables. The logistic function provides the probability of an event belonging to a category and takes the form:

$$p(X) = \frac{1}{1 + e^{-(\boldsymbol{\beta} \cdot \boldsymbol{X})}}$$

where X is a vector representing a sample and $\boldsymbol{\beta}$ is a vector of model coefficients to be learned by the model [27].

Logistic regression classifiers need to be trained before we can use them. Training a logistic regressor means feeding it with training data so it can learn $\boldsymbol{\beta}$, an internal representation that minimizes the training error. This training set has the same form as the data we intend to classify in the future, plus a label; this is the solution, or the piece of data that we want the classifier to predict. After training, the logistic regressor can model the expected output as a function of the input. Once our classifier is trained, we can feed it with non-labeled data and obtain predictions of the label. This kind of ML is called supervised learning.

**4. Proposed model**

To achieve our goals, we propose a model based on CRISP-DM [28]. This model is shown in Fig.1

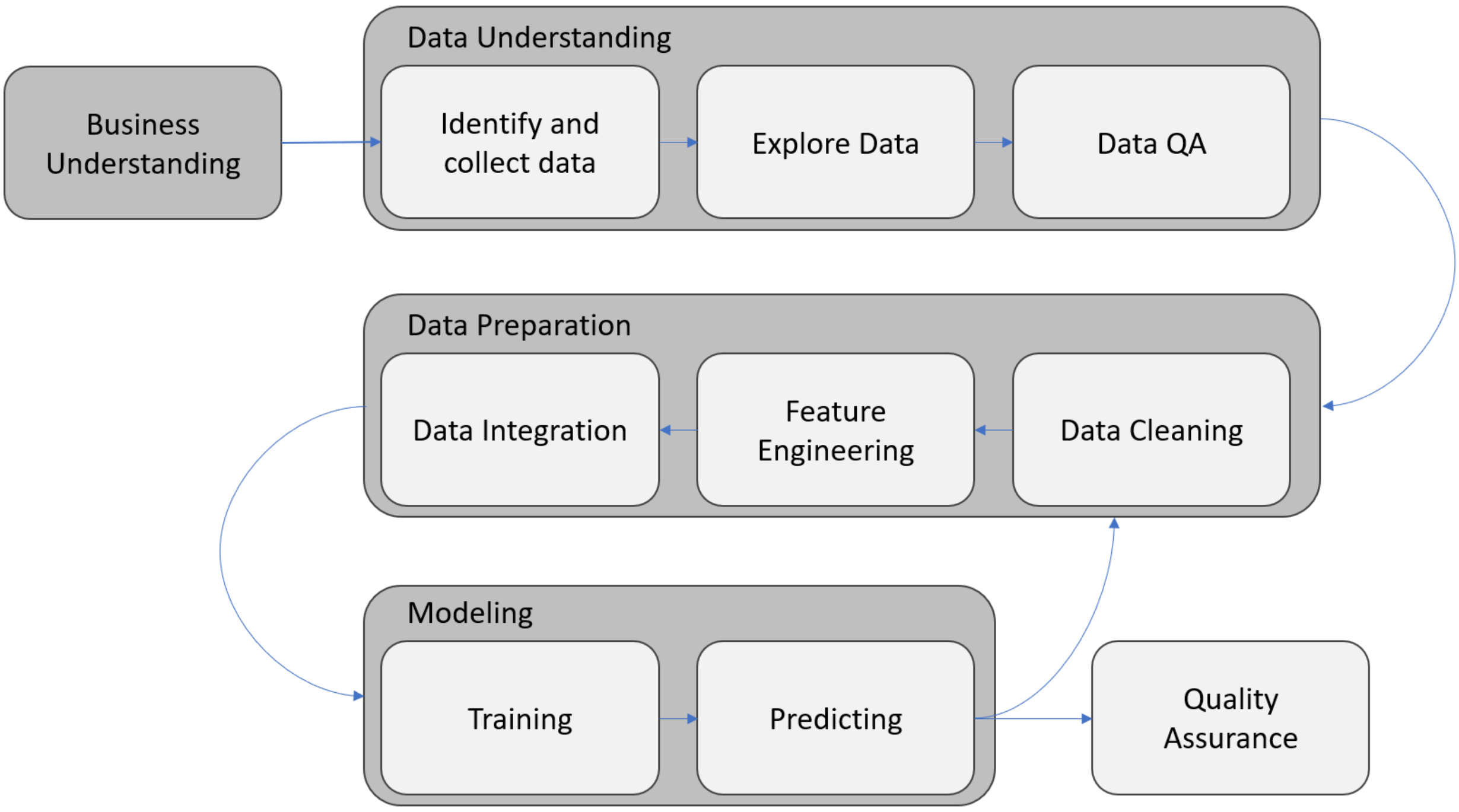

**Figure 1 – Proposed model**

Below, we explain each of the phases and apply them to a real case. For this real-world application, we use the general data architecture based on [29] and [30]. This architecture allows us to apply our model in a scalable big data context.

Applying those principles to our project, our data flow is as shown in Fig.2.

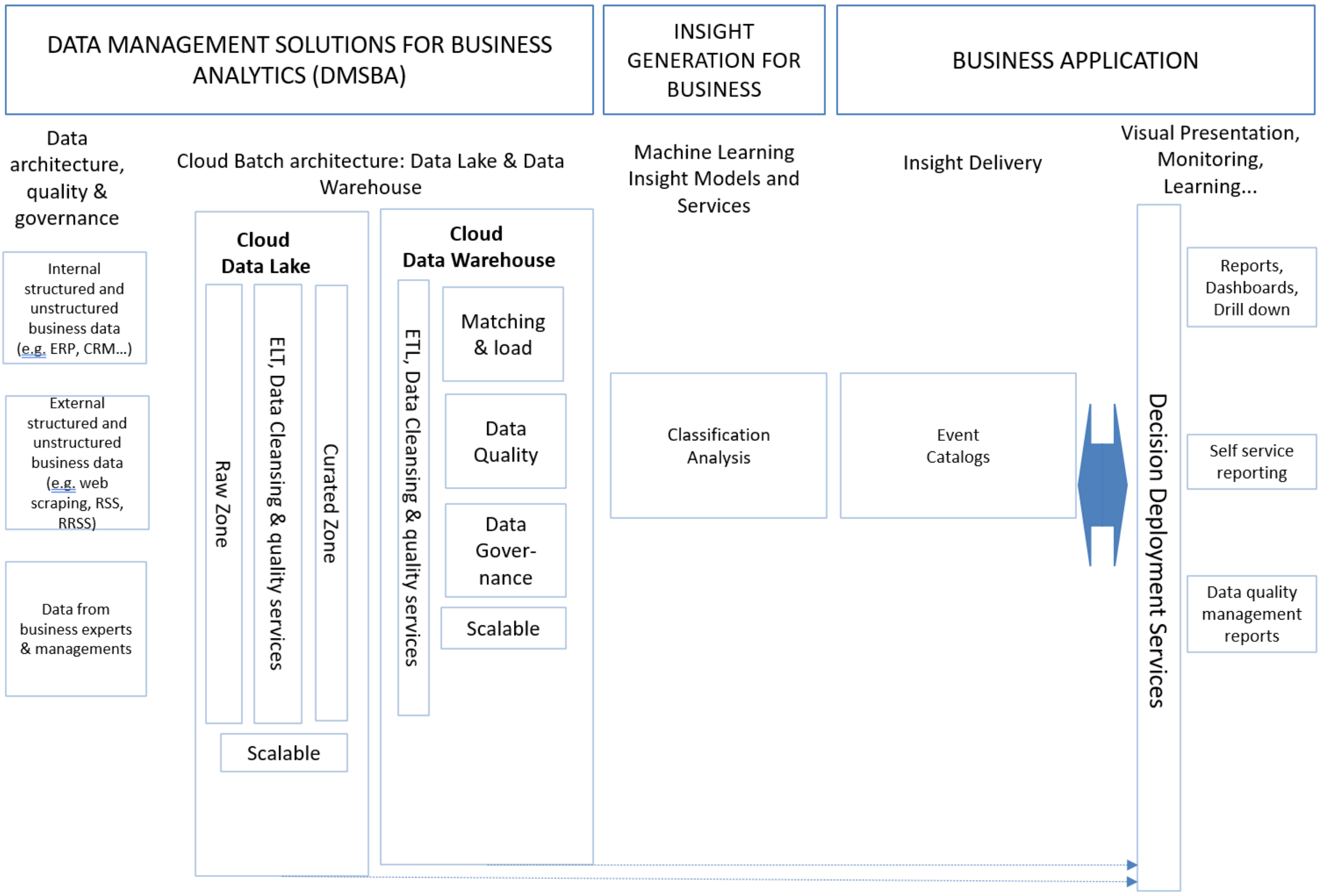

**Figure 2 – Data-driven architecture**

The components of the conceptual architecture are as follows:

- Data management solutions for business analytics (DMSBA). Event-related data are collected using APIs, RSS feeds and web scraping, and stored raw in a NoSQL database. Raw data are then validated, cleaned, normalized, enriched, and related to existing data. After this curating process, the data are stored in a relational database.
- Insight generation for business. At this point we use ML to further enrich data and gain knowledge that we can use in business applications.
- Business applications: The knowledge generated in the previous layer is used to drive different applications such as the creation of event catalogs, event relevance assignation, inspirational travel services, traffic prediction and transportation, hotel occupancy, city planning or sustainable tourism.

*4.1 Business Understanding*

The objective is to obtain a model that can be used to predict the taxonomic category of an event based on its title and description.

Our system must be able to take a piece of text and classify that text belonging to an event given a fixed set of taxonomic categories.

A taxonomy is a specific classification scheme that expresses the overall similarity between organisms, entities and/or things in a hierarchical fashion [29]. Items sharing similarities are first grouped into populations. According to [7], a product hierarchy extends from basic needs to items that satisfy those needs. In our case, part of the business understanding step is understanding the need to classify events in a taxonomy that is useful to the sector.

This is a very interesting feature for tourism industry players such as airlines. In our example, a particular airline flies to 20+ countries and 50+ cities. This airline wants to provide an inspirational service that allows users to browse the tourist events in those cities, and hopefully encourages them to buy a ticket if the event catalog impresses them enough. The catalog should be filterable by type of event; hence, all events should belong to at least one taxonomic category. This airline cannot use an aggregated catalog since events collected from hundreds of different sources and happening in 20+ different countries do not share the same classification criteria, if any have been assigned. Instead, they will use a hierarchical taxonomy to classify the events that they present to their clients.

An example of how five events may look in a simplified catalog is shown in Fig. 3.

| | title | description | taxonomy | starts | ends | longitude | latitude | city |
|---|---|---|---|---|---|---|---|---|
| 0 | London Dungeon LATES with Cocktail | Lates mashes up theatre, special effects and i... | other interesting events | 2017-03-29T19:00:00+00:00 | 2019-11-22T22:00:00+00:00 | 51.502820 | -0.119252 | London |
| 1 | Drumchapel & West Winterfest Fireworks | Don't miss Drumchapel's annual fireworks extra... | other interesting events | 2019-11-05T00:00:00+00:00 | 2019-11-05T20:59:59+00:00 | 55.901126 | -4.373647 | Glasgow |
| 2 | MEGALAND 2019 | MEGALAND 2019 in the Simón Bolivar Park. Live ... | music | 2019-11-30T13:00:18+00:00 | 2019-12-01T02:00:18+00:00 | 4.659293 | -74.093524 | Bogotá |
| 3 | Extravaganza | It's Extravaganza time at Ferrymead Heritage P... | other interesting events | 2019-10-26T00:00:00+00:00 | 2019-10-27T23:00:00+00:00 | -43.567162 | 172.702541 | Christchurch |
| 4 | A Nightmare on Duddell's Street | Disco Bao is back, this time with a freaky twi... | music | 2019-10-26T22:00:00+00:00 | 2019-10-26T23:59:59+00:00 | 22.280244 | 114.157230 | Hong Kong |

**Figure 3 – Example of five events in a simplified catalog**

*4.2 Data Understanding*

Data understanding can be divided into four tasks:

- Determine what data we need
- Collect data
- Explore data

- Quality Assurance (QA)

First, we had to determine what data we needed; for our problem, it was a set of records representing events. Since our goal is to predict the taxonomic category of an event based on its title and description, we had to collect title and description as independent variables and assign a label with an identifier of its taxonomic category as a dependent variable.

The next step was to collect the data. To do so, we created a series of data collectors able to ingest large amounts of data. These data collectors fetch data from very different sources such as APIs, RSS feeds, data aggregators and more than 1000 websites from more than 30 countries in 22 different languages. Since the raw data from those sources are not evenly structured, we used a NoSQL database to store the results. We built this process iteratively for six years and collected millions of events with rich data that go beyond title and description.

Data exploration was an important part of the work. Data exploration showed that many events we were collecting could be aggregated by language, country, region, venue, date, etc. Since this collection continued over a span of six years, we could modify the system to expedite some sources, build new ones if needed and react to collection results.

After collecting the data, we observed that our records were unbalanced due to the heterogeneous nature of the events catalog. In Fig. 4 we can see that that we had twice as many events with id 0 (music) as some of the other taxonomic categories in our set.

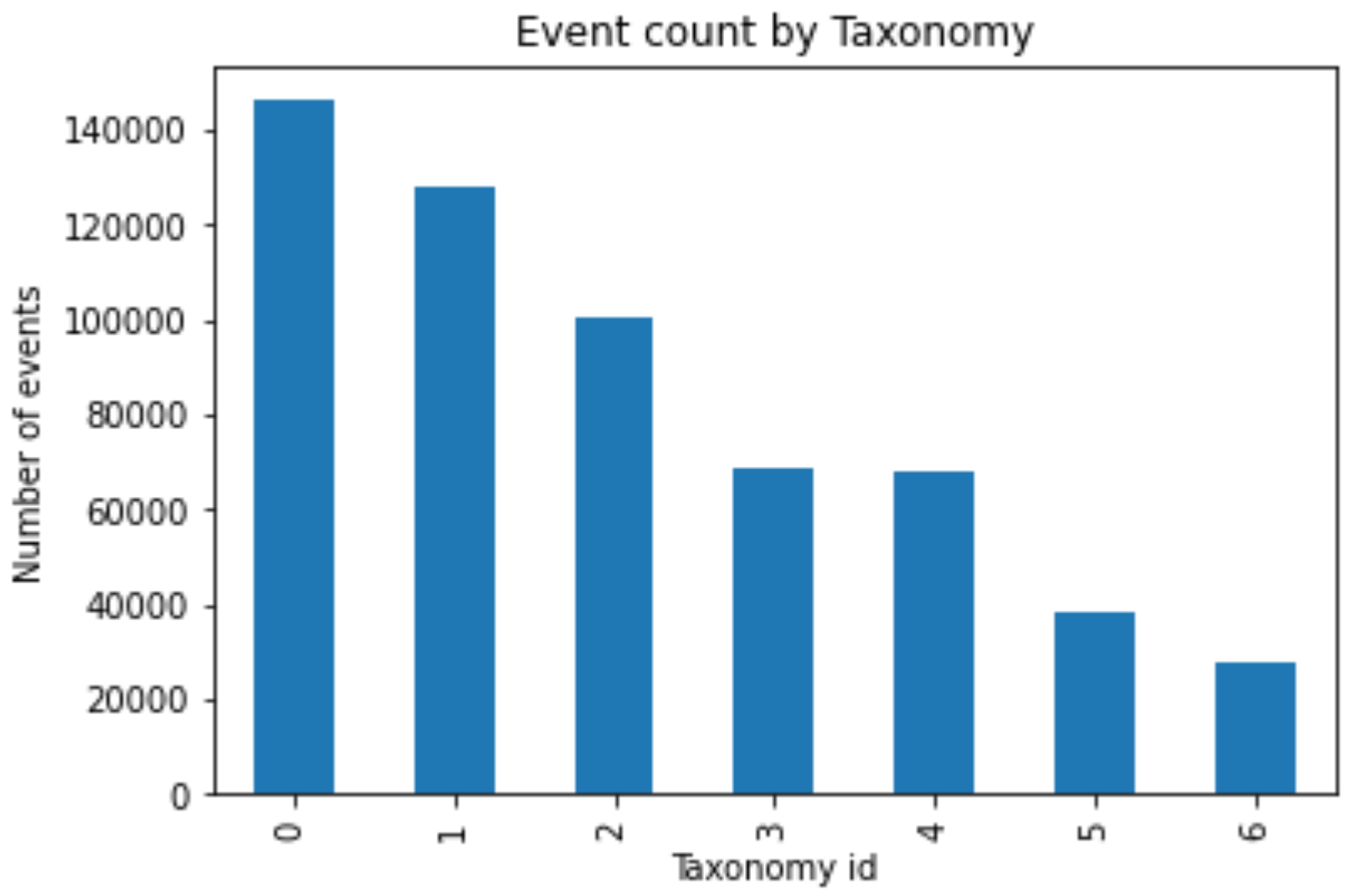

**Figure 4 – Count of first level taxonomic categories in our set.**

QA was performed periodically to check that information was consistent and we were collecting quality data. There were some automatic measures in place to ensure only valid events were ingested.

*4.3 Data Preparation*

Data preparation can be divided into three tasks:

- Data cleaning
- Feature engineering
- Data integration

Data cleaning is necessary when collecting a high volume of data directly from the web. The process must deal with multilingual texts and issues such as HTML contamination or formatting issues. Most of these issues are solved by erasing common HTML symbols or expressions, non-ASCII characters, email addresses or most punctuation characters. In Fig. 5 we show what some events looked like before cleaning them, with HTML tags, JSON formatting, punctuation such as bullet points, quotes, non-ASCII characters, etc.

```
['Works by Jean-Philippe Rameau, François Couperin and Marin Marais \\u003cbr /\\u003e \\u003cbr ,
['"On January 2, 2019 I will introduce Nivuru at the Teatro Biondo.When the new year and the reger
['Underwater Photography Course Base LevelPROGRAMMHall 1 (classroom): • Beginning 09: 30 • Underst
['{ "Type": "Video", "object": { "title": "Mika_Tour_Italy_Low_Res_Fades-2", "description": "", "t
['TIA DE Carlos "Nicolás Olivari\'s free version of Brandon Thomas\'s famous play was premiered at
['#BeSocial & let it grow, grow, grow all November long in support of the Movember Movement!Movemt
['<p style="text-align:center"><img src="web width="550" height="743" border=" 0" hspace="0" vspac
```

**Figure 5 – Event name and description before cleaning.**

Another important data transformation is translation. In our case, if an original event was not in English, we had to automatically translate it since the BERT model that we used is optimized for the English language. Data integration happened naturally in our system since we matched venues, events, performers, languages, countries, and regions in a relational database. A crucial step in this process was to label events with a category and a second level category, for which we used different techniques:

- Some of the events came from sites that only dealt with one taxonomic category of events and hence they were automatically labeled. For example, if one of our second level taxonomic categories was football and we collected events from a source that only lists football-related events, we labeled those events as football when collecting them.
- Some of the events came from reputed sites with some taxonomic categories that exactly matched some of ours. In such cases we were able to label events using a simple data mapping.
- A large amount of data was labeled manually by data analysts.

Our list of first level taxonomic categories is shown in Fig. 6.

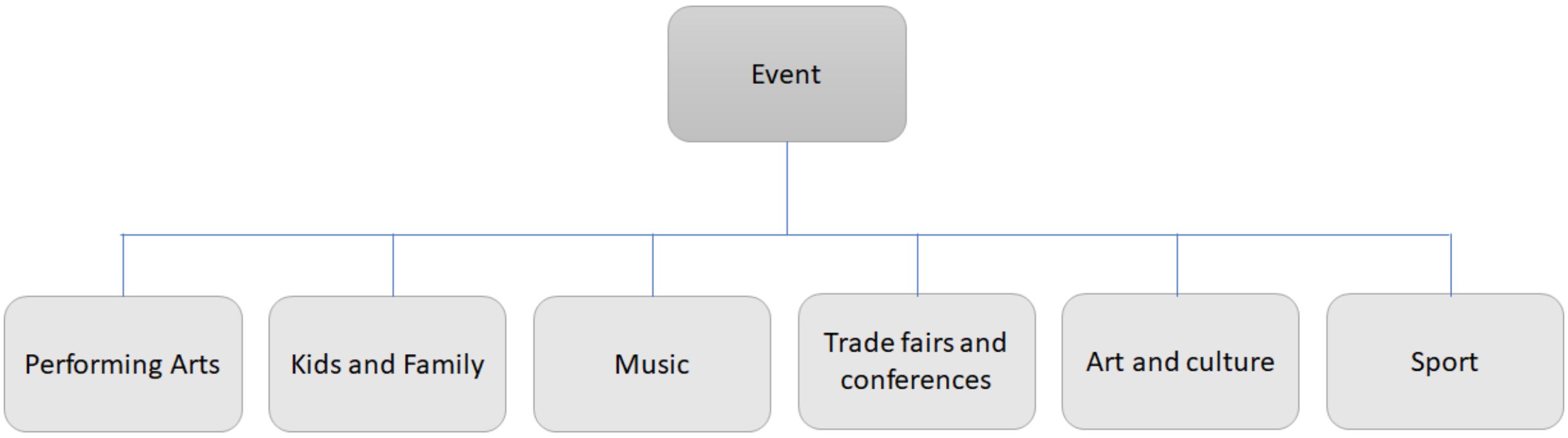

**Figure 6 – Event taxonomy tree**

At this point, after completing these tasks, we ended up with 700,000 events with a clean title and description and labeled according to our taxonomy structure.

Since we wanted to feed BERT with only one text string per event and the maximum number of words that we can use without further modifications to the model is 510, we concatenated title, a blank character, and description and trimmed it to 510 words. This will affect some of our events, but it does not seem to affect classification accuracy. In Fig. 7 a probability distribution of our dataset is shown. We can see that the most common cases are events with short texts.

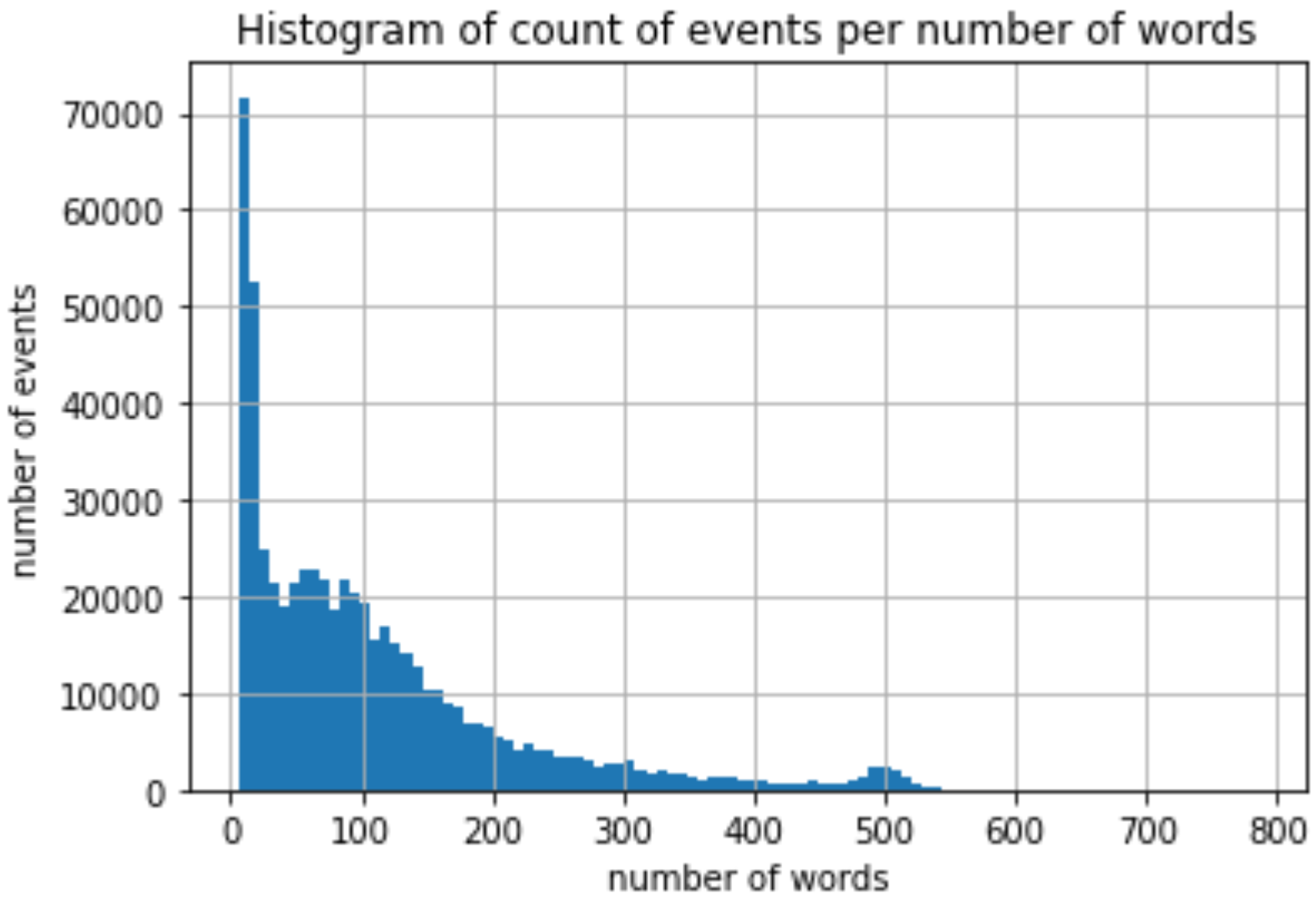

**Figure 7. Probability distribution of events by number of tokens.**

In our example of how an airline could use this catalog, the process is as follows: the airline would select the sources of events and would apply some restrictions such as event blocking by keywords or other considerations. Then we would generate the catalog using web scrapers, RSS feeds, etc. as described in Section 4.2. If the source material was not in English, we would translate the title and description. We show in Fig. 8 how the collected title and description look after translation and before cleaning.

| | title | description |
|---|---|---|
| **5089** | Wall Sky Lounge at Friedrichstadt-Palast Berli... | Experience the breathtaking show VIVID in the ... |
| **8748** | The performance of "Bells and spells� at the T... | When, in the 1970s, the movement for the prote... |
| **9082** | Ã', Spain Spectacular Show | A show based on music, dance and audiovisual, ... |
| **10761** | Those of Barranco | A comedy directed by Toto Castiñeiras during h... |
| **10764** | Cirque Du Soleil - ovo | The prices of show tickets are dynamic and can... |
| **12754** | Long form intensive with direct feedback - ben... | Through various theatrical techniques and exer... |
| **14211** | International Workshop of Clown with Anton Val... | The ImaginoTeatro School, has the honor of pre... |
| **24814** | The Beatles: Love (Teatro Melico Salazar) | We want to revive the Beatles scene in our cou... |
| **26442** | An Exciting Journey Through the World of Kooza | From Cirque du Soleil\nscreening of the docume... |
| **33982** | Cirque du Soleil in Guatemala | Cirque du Soleil will arrive in Guatemala, ver... |

**Figure 8. Title and description of events before cleaning**

This catalog is then cleaned, and title and description are concatenated. We can see this process in Fig. 9

| | title | description |
|---|---|---|
| 0 | London Dungeon LATES with Cocktail | Lates mashes up theatre, special effects and i... |
| 1 | Drumchapel & West Winterfest Fireworks | Don't miss Drumchapel's annual fireworks extra... |
| 2 | MEGALAND 2019 | MEGALAND 2019 in the Simón Bolivar Park. Live ... |
| 3 | Extravaganza | It's Extravaganza time at Ferrymead Heritage P... |
| 4 | A Nightmare on Duddell's Street | Disco Bao is back, this time with a freaky Twi... |

| | title_description |
|---|---|
| 0 | London Dungeon LATES with Cocktail. Lates mash... |
| 1 | Drumchapel & West Winterfest Fireworks. Don't ... |
| 2 | MEGALAND 2019. MEGALAND 2019 in the Simón Boli... |
| 3 | Extravaganza. It's Extravaganza time at Ferrym... |
| 4 | A Nightmare on Duddell's Street. Disco Bao is ... |

**Figure 9. Title and description after cleaning and after concatenation**

### *4.4 Modeling*

After finishing the data preparation, the next step was to build a model that allowed us to classify events.

#### *4.4.1 Training*

To train our model, we had to train a classifier able to predict the taxonomic category of the event. Then we needed to train more classifiers capable of determining the second level taxonomic category of the event. We trained one of those second level classifiers for each taxonomy. The internal architecture of each classifier was the same; to build different classifiers we fed the training process with different data.

The overall training process of our classifier entailed sending events to be predicted (forward pass), comparing prediction with ground truth to get a classification error and sending that error back so the network could minimize it (backpropagation). The network in our case was formed by hundreds of layers provided by the BERT model and one layer for the logistic classifier.

In the forward pass, we sent an event to the classifier. The input of each classifier when training included the event text, a mask indicating the length of the text and the label of the event. Our text was then sent to a tokenizer that produced the tokens that BERT can consume. The output of BERT was a 768-length vector with a representation of the input text. That vector was fed into a multiclass logistic regression classifier that produced an array with probabilities, one per taxonomic category. Finally, the softmax activation function was applied to the output vector, which picked the largest probability and returned the most probable taxonomic category.

Having obtained a prediction, the system sent the classification error back to the network in a process called backpropagation. Then both the logistic regression classifier and BERT learned by trying to minimize that error.

We repeated this forward and back propagation with all the events in our training set, and then repeated the entire operation for 4 epochs. The output of this process was a model that we can store and retrieve to classify events.

*4.4.2 Predicting*

To classify an event, we sent it to the BERT-based classifier that we trained in the previous step and thus determined its taxonomic category, e.g. Sport.

A BERT-based classifier receives a clean string of text up to 510 characters in length which is then tokenized using the appropriate tokenizer. These tokens are fed into the BERT model that produces a vector representation of the event text. That vector goes through a trained logistic regression and a softmax activation function that returns the category or second level category.

We can see the workflow of our classifier in Fig. 10.

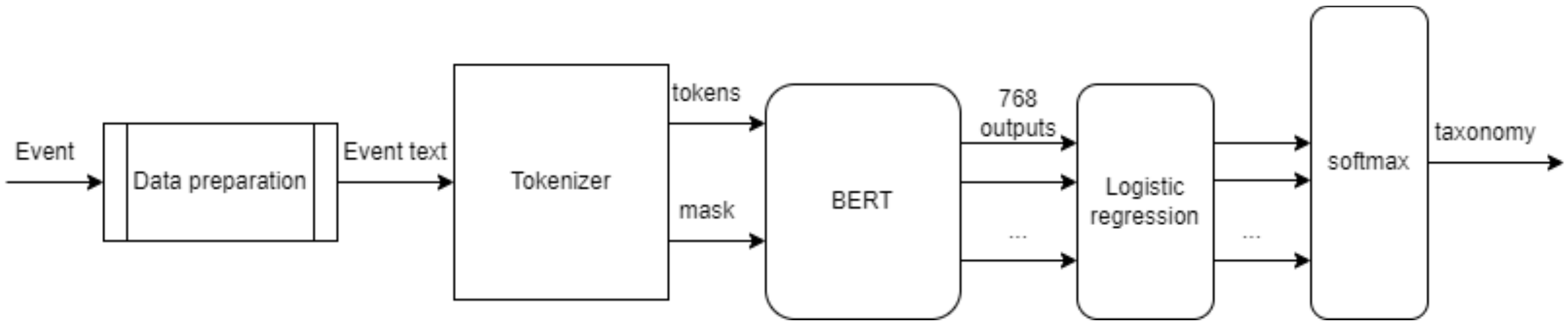

**Figure 10. Data flow for each taxonomy**

In our use case, once we have a clean concatenated dataset, the airline has all the events in their target cities ready to be classified. We send this text to our model that assigns a taxonomic category. Following the example shown in Fig. 7, we see an event with the title "Cirque Du Soleil - ovo". Our classifier will assign the taxonomy "Performing Arts" to the event.

*4.5 Evaluation*

We evaluated our model using 173,184 events. These 173,184 events were stratified since the original training dataset was unbalanced. For this dataset we obtained an accuracy of 0.87.

To obtain that accuracy we generated a table with the logit odds of each sentence, which represented the likelihood of each event belonging to each taxonomic category. We added a column with the taxonomy id with the highest odds and considered that our prediction. This is shown in Fig. 11, where the odds are shown in the numerical columns.

| | 0 | 1 | 2 | 3 | 4 | 5 | 6 | pred | ground | sentence |
|---|---|---|---|---|---|---|---|---|---|---|
| 0 | 6.590662 | -0.590513 | -2.354302 | -4.126739 | -2.676668 | -3.413306 | -4.567245 | 0 | 0 | Kurt Rosenwinkel Marni Jazz FestivalStandards ... |
| 1 | -2.390227 | -2.297226 | 5.807707 | -4.175766 | -1.344302 | -2.657254 | -1.747721 | 2 | 2 | THE SECRET GARDEN OF THE CAPE Exhibition of th... |
| 2 | 3.557938 | -0.077841 | -3.069335 | -3.932930 | 0.489278 | -0.425255 | -3.674125 | 0 | 0 | What Is Music Festival Views Price free, free ... |
| 3 | -1.243803 | 6.211366 | -2.031531 | -4.946661 | -2.636188 | -4.949661 | -3.065950 | 1 | 1 | Rigoletto Komische Oper Berlin A comic opera t... |
| 4 | 0.185281 | -0.412477 | 4.460521 | -3.982399 | -1.978161 | -2.249258 | -4.697901 | 2 | 2 | Persephonium Andreana Dobreva Andreana Dobreva... |
| ... | ... | ... | ... | ... | ... | ... | ... | ... | ... | ... |
| 173170 | -2.992995 | -2.773311 | 6.163981 | -3.372595 | -2.207108 | -1.622462 | -2.266437 | 2 | 2 | Exhibition at Art Corner Gallery Art Corner Ga... |
| 173171 | 4.184312 | -1.935361 | -4.027596 | -3.060884 | -0.357280 | -3.512146 | 1.069656 | 0 | 0 | Lucia concert with Adolf Music School Stockhol... |
| 173172 | -0.811240 | -3.428475 | -4.322423 | -2.099226 | 5.157854 | -2.065966 | 0.581620 | 4 | 4 | Chipping Norton Town Festival A day of music, ... |
| 173173 | -4.189793 | -4.220690 | -2.661265 | -2.548733 | 5.360709 | 2.235356 | -1.612044 | 4 | 4 | Christmas market in Christmas market in As eve... |
| 173174 | -4.138300 | -2.972013 | -1.283450 | -3.456191 | 0.634896 | 5.016858 | -0.698809 | 5 | 5 | Discussion Creation and technique in live codi... |

**Figure 11. Logit odds and predictions for each sentence in our test set**

Table 2 contains a more detailed report of our test metrics including metrics per category.

**Table 2. Classification report**

| | **Precision** | **Recall** | **F1-score** | **Support** |
|---|---|---|---|---|
| **Category 0 (music)** | 0.87 | 0.92 | 0.90 | 43839 |
| **Category 1 (performing arts)** | 0.88 | 0.84 | 0.86 | 38372 |
| **Category 2 (art and culture)** | 0.88 | 0.87 | 0.88 | 30088 |
| **Category 3 (sports)** | 0.97 | 0.97 | 0.97 | 20546 |
| **Category 4 (other events)** | 0.81 | 0.80 | 0.80 | 20337 |
| **Category 5 (trade fairs and conferences)** | 0.84 | 0.85 | 0.84 | 11567 |
| **Category 6** | 0.76 | 0.78 | 0.77 | 8426 |

**(kids and family)**

| | | | | |
|---|---|---|---|---|
| accuracy | | | 0.87 | 173175 |
| macro avg | 0.86 | 0.86 | 0.86 | 173175 |
| weighted avg | 0.87 | 0.87 | 0.87 | 173175 |

The best classified taxonomic category was Category 3 (sports). This may be because sport events are more consistently worded across sources. Also, sport events have a smaller overlap with other events. It is reasonable to think that categories 1, 2 and 4 ('performing arts', art and culture' and 'other events') sometimes overlap. Taxonomy id 6, 'kids & family', showed the worst performance, possibly because it was the smallest set, and the model could not learn it as well as the other taxonomic categories. Another reason may be a big overlap with other categories. For example, a concert for children may confuse our model since that event could fall under 'music' or under 'kids & family'.

We can dive into this overlap problem by observing a confusion matrix and the normalized confusion matrix, as shown in Fig. 12.

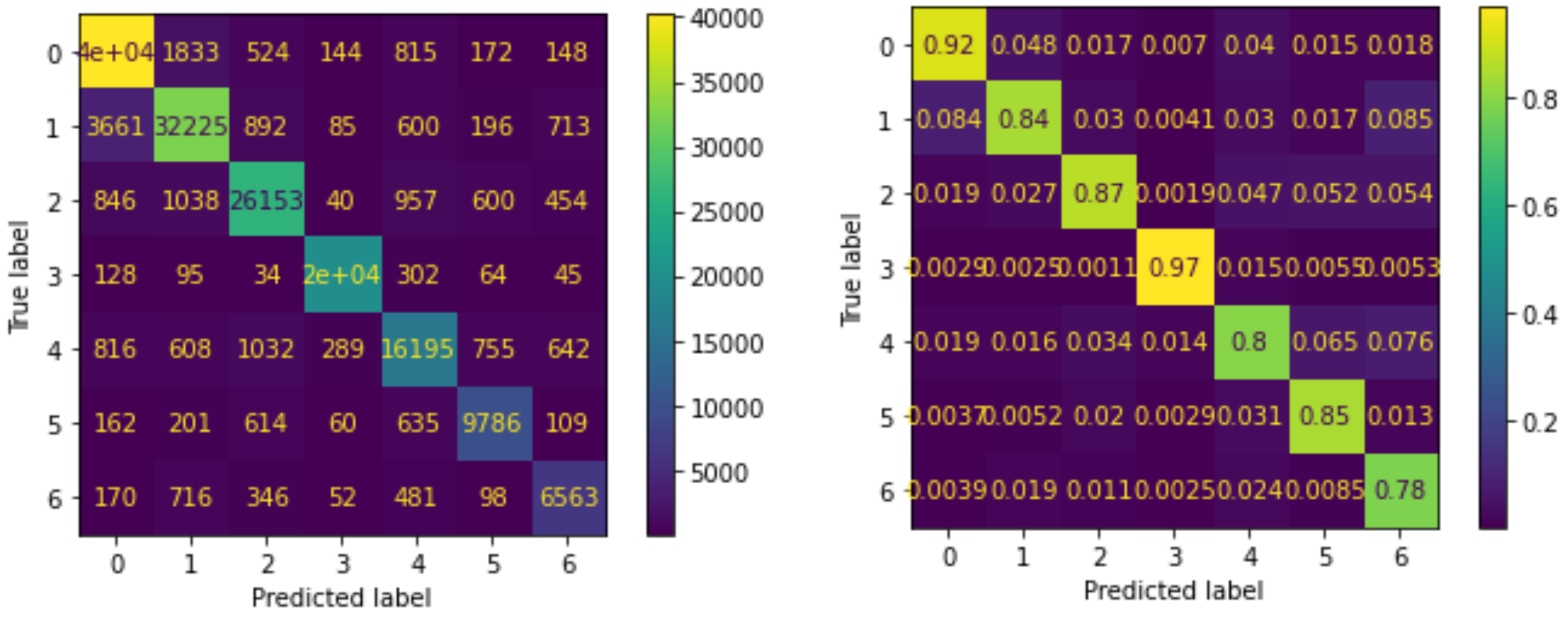

**Figure 12. Confusion matrix and normalized confusion matrix**

If we focus on the worst performing taxonomic category, Taxonomy id 6, we can see that it is often confused with Taxonomy id 4, as 2.4% of 'kids & family' events are wrongly classified as 'other events'.

In our use case, the airline could classify events using different techniques. BERT is expensive compared to other methods, so we use it only when necessary. Other methods include:

- Assignation of taxonomic category if the source is trustworthy and has the same category. For example, we could assign the taxonomic category 'music' to all events from a source that only delivers music.
- Assignation of taxonomic category if indicated by other data. For example, our airline collects more data than title and description of events and can use them to help classification. It may classify all events as 'trade fairs and conferences' if they are held in a venue that only hosts that kind of event.

We obtained an accuracy of 0.87 which is quite high for our purposes. Our classification approach involves assigning taxonomic categories to events based on the site that published them, the venue, etc. We only use NLP techniques for events that cannot be classified using cheaper techniques. This means that our general catalog has a much higher accuracy than 0.87 and the airline can safely use it to interact with its clients.

## 5. Discussion

We built a hierarchical NLP model that assigns a fixed set of taxonomic categories to tourist events. This model allows us to normalize events so we can build a catalog that is useful to tourism industry players such as airlines, travel agencies or hotel chains.

With a normalized catalog we can use consistent search filters independently of the source, origin language or geographical location. It is especially useful for travelers during the inspirational phase of their trip-planning. We obtained an accuracy of 0.87; thus, combined with other classification techniques, our model can produce a quality catalog for the tourism sector.

Concluding with our use case, an airline can use this catalog to know what events are happening in its regions of interest and can provide services where users can filter events by various different criteria. The airline can build its own catalog based on the general one; for example, if it takes a complete catalog showing all published events from our sources, the airline may choose not to show some of the taxonomic categories. Some interested parties may not want to show the second level taxonomic category "alcohol and beverages". Once we have a catalog with all target events classified and filtered, the airline may use it in different ways:

- It could build an application to show some taxonomic categories across its destinations, to help travelers in their inspirational phase
- It could filter some interesting events for a specific destination if it wants to promote that route
- It could tailor and personalize the catalog for its clients if it knows their preferences.

## *6. Conclusions and Future Work*

This work addresses a significant challenge in the tourism industry: the lack of a unified and normalized catalog of events. Most of the related literature proposes taxonomies or typologies for events, but does not describe how to apply them or how to assign them to existing events. The novelty of our research lies in the proposal of a practical solution to automatically classify events using a fixed set of taxonomic categories, shifting the burden from event organizers and listing sites to event aggregators. To our knowledge, our study is the first to specifically tackle this problem.

We built a BERT-based classifier which was able to assign taxonomic categories to an event catalog using only titles and descriptions of events. We trained this classifier with 700,000 events and tested it with 173,175 labeled events. In the future, we can experiment with other variants of BERT. We can also allow events to belong to multiple categories to prevent issues with overlap, or we can generate taxonomic categories that are easier to separate completely.

From the economic point of view, companies need normalized event catalogs for multiple reasons. In this study, we use the example of an airline providing a normalized catalog for events in many cities, in many different languages, and from very different sources, to help travelers in the inspirational phase. A future line of work can focus on other analyses, such as the prediction of venue occupation, hotel occupation or transport usage.